          [2001/04/25 1.1 (PWD)]
\documentclass[a4paper,twocolumn]{esapub2005} % European paper
\usepackage{times}
\usepackage{natbib}
\usepackage{graphicx}

\title{Molecular Hydrogen bubbles formation on thin vacuum deposited Aluminum layers after proton irradiation.}
\author{Maciej Sznajder}
\affil{DLR Institute for Space Systems, University of Bremen, Robert Hooke Str. 7, 28359 Bremen, Germany}
\author{Ulrich Geppert}
\affil{DLR Institute for Space Systems, System Conditioning, Robert Hooke Str. 7, 28359 Bremen, Germany}
\affil{Kepler Institute of Astronomy, University of Zielona G\'{o}ra, Lubuska 2, 65-265 Zielona G\'{o}ra, Poland}

\begin{document}


\maketitle

\begin{abstract}
Metals are the most common materials used in space technology. Metal structures, while used in space, are subjected to the full spectrum of the electromagnetic radiation together with particle irradiation. Hence, they undergo degradation. Future space missions are planned to proceed in the interplanetary space, where the protons of the solar wind play a very destructive role on metallic surfaces. Unfortunately, their real degradation behavior is to a great extent unknown.

Our aim is to predict materials' behavior in such a destructive environment. Therefore both, theoretical and experimental studies are performed at the German Aerospace Center (DLR) in Bremen, Germany. 

Here, we report the theoretical results of those studies. We examine the process of $\rm{H_2}$-bubble formation on metallic surfaces. $\rm{H_2}$-bubbles are metal caps filled with Hydrogen molecular gas resulting from recombination processes of the metal free electrons and the solar protons. A thermodynamic model of the bubble growth is presented. Our model predicts e.g. the velocity of that growth and the reflectivity of foils populated by bubbles.   

Formation of bubbles irreversibly changes the surface quality of irradiated metals. Thin metallic films are especially sensitive for such degradation processes. They are used e.g. in the solar sail propulsion technology. The efficiency of that technology depends on the thermo-optical properties of the sail materials. Therefore, bubble formation processes have to be taken into account for the planning of long-term solar sail missions.
\end{abstract}

%-----------------------
\section{Introduction}

Vacuum deposited Aluminum layers on thin polyimide films are commonly used composite materials in space technology. Unfortunately, space environmental conditions cause changes of their mechanical and thermo-optical properties. Therefore studies, both theoretical and experimental, allow to choose their properties, e.g. layer thickness, which may well fit to a given space mission.

The influence of the interplanetary space environment onto thin metallic films is to a great extend unknown. Here, a candidate for aging process is proposed: formation of molecular Hydrogen bubbles onto metallic surfaces resulting from recombination processes of solar protons and the metal free electrons. Degradation of structural properties of solids caused by Hydrogen (referred to as embrittlement) plays a fundamental role in materials physics \citep{lu}. Bubble formation is one of the four general processes of embrittlement \citep{lu}.  

Here a thermodynamic model of molecular Hydrogen bubbles formation under space conditions is presented. The model input parameters are: energy and flux of solar protons, type, and temperature of the irradiated metal. The diffusivity of H in the metal lattice was taken into account, as well as back scattering effect (BS) of the solar protons irradiating the target. The model output is the velocity of bubble radius growth, the maximum possible bubble radius, and, for a given bubble density and average bubble radius, the reduction factor of the reflectivity with respect to its ideal value. 

The paper is organized as follows. In Section \ref{blistering_space}, general principles and conditions for the $\rm{H_2}$-bubble formation are given. In Section \ref{th_model}, the thermodynamic model of bubble formation is introduced \citep{szna}. Then, the effect of bubble growth onto the specular reflectivity is discussed in Subsection \ref{spec_reflectivity}. In Section \ref{results} the experimental results as well as the validation of the model are presented. Finally, in Section \ref{conclusions}, the conclusions are drawn.  

%-----------------------------------------------------------------------------------
\section{Formation of molecular hydrogen bubbles under space conditions} \label{blistering_space}

Formation of molecular Hydrogen bubbles depends on many physical parameters, for instance: the type of the irradiated material, the proton energy, the proton flux, the temperature of the target, the crystallographic orientation of the irradiated surface as well as on the impurities and defects in the sample. It is known from terrestrial laboratory experiments that the minimum dose of protons above which the process occurs is $~10^{16}$ $\rm{H^+}$ $\rm{cm^{-2}}$ \citep[e.g.][]{milacek}. The temperature range in which bubbles were observed is between 280 and 570 K \citep[e.g.][]{daniels, milacek}. 

The procedure used to estimate the critical temperature ($570$ K) above which the process of bubble formation stops due to the bubble cracking mechanism was as follows. The Aluminum target was irradiated by a flux of protons at room temperature. When irradiation of the sample was stopped, the probe was heated up to higher temperatures. A significant increase of both, the surface density and sizes of the bubbles has been observed until the critical temperature of bubble cracking was reached. That procedure, used by the authors, allows to capture more Hydrogen by the vacancies since during the irradiation, and at room temperature the vacancies will collect more Hydrogen than at elevated temperatures. Also a diffusion of Hydrogen in Aluminum at room temperature is much lower than at temperatures reaching $\sim570$ K \citep{lind}. In space a probe is bombarded by solar protons at temperatures which are related to their orbit. Therefore, the procedure presented by \citep{daniels,milacek} does not match the bubble formation mechanism under real space conditions.

Growth of molecular Hydrogen bubbles will be possible in the interplanetary space if the criterion of the minimum dose of protons is fulfilled. The temperature of the sample has to be high enough to start the bubble formation, but not too high to lose Hydrogen much too rapidly due to the high diffusivity of Hydrogen in metals. 

\begin{figure}
  \centering
    \includegraphics[width=0.5\textwidth]{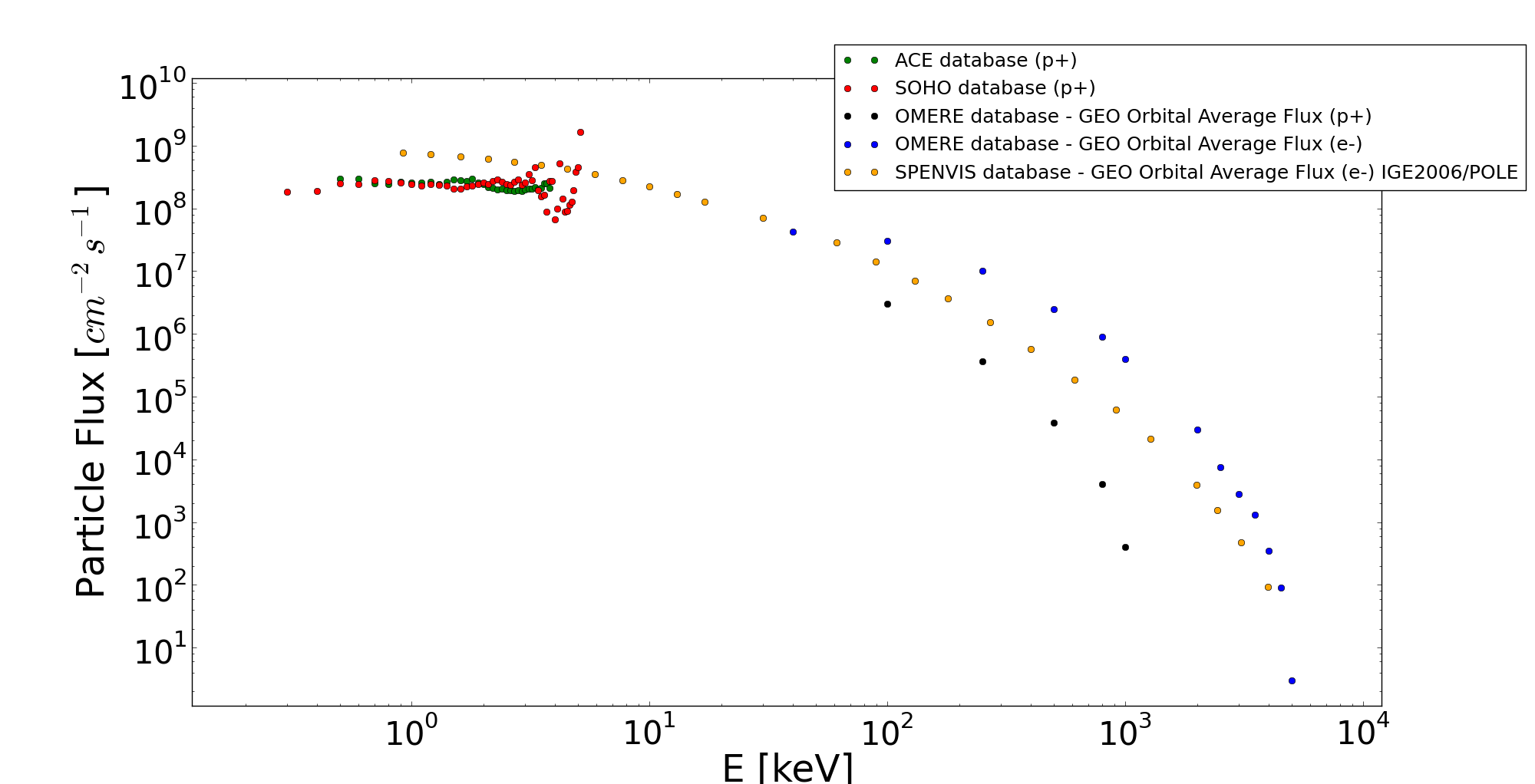}
  \caption{Flux of solar protons and electrons as a function of energy. Data are taken from the SOHO, ACE, OMERE, and SPENVIS databases.}
  \label{p_flux}
\end{figure}

Fig. \ref{p_flux} shows solar proton and electron fluxes at 1 AU distance from the Sun. Proton fluxes are calculated by use of the data collected by the SOHO (since 1995) and the ACE (since 1997) satellites. The OMERE as well as the SPENVIS databases were also considered. 

When a probe is irradiated in space, it collects incident ions from a wide energy range. The range depends on the type and the thickness of an irradiated material. The thiner the target material, the less ions stuck in it. Therefore, there must exist a critical energy of incident ions ($E_{\rm{C}}$) above which they pass through the material. Hence, the integrated proton flux over the energies is:

\begin{equation}
  I_{\rm{E}} = \sum_{E_{\rm{min}}}^{E_{\rm{c}}} I(E),
\end{equation}  

\noindent where $E_{\rm{min}}$ is the ion's lowest energy recorded by the satellite's detector system. The $I_{\rm{E}}$ values are presented in the Table \ref{integrated_proton_flux}. To calculate the fluxes the ACE database was used.  

\begin{table}[!ht]
\centering 
\caption{Integrated proton fluxes over the energies for $1$ AU distance orbit from the Sun.}
\begin{tabular}{cc}
\hline
$E_{\rm{c}}$ [keV] & $I_{\rm{E}} \times 10^{13}$ [$\rm{p^+}$$\rm{cm^{-2}}$$\rm{s^{-1}}$]  \\
\hline
1.0 & 0.44 \\
1.5 & 0.68 \\ 
2.0 & 0.91 \\
2.5 & 1.06 \\
3.0 & 1.12 \\
4.0 & 1.14 \\
5.0 & 1.15 \\
9.0 & 1.15 \\
\hline
\end{tabular}
\label{integrated_proton_flux}
\end{table}
To estimate the flux of solar protons $I_{\rm{r}}$ at distance $r$ from the Sun, the following relation can be used:

\begin{equation} \label{flux_r}
  4\pi (1\rm{AU})^2 \times I_{\rm{1 \ AU}} = 4\pi r^2 \times I_{\rm{r}}.
\end{equation}

Under the simplifying assumption that the Sun generates only mono-energetic 5 keV protons, the criterion of minimum dose of protons will be fulfilled after 116 days for 1 AU distance orbit from the Sun. Obviously, taking into account proton fluxes from the whole energy range, the criterion will be fulfilled much earlier.

The temperature of a foil placed in a given distance $d$ from the Sun can be calculated by:

\begin{equation}
  T = \left( \frac{A_{\rm{a}}}{A_{\rm{e}}} \frac{\alpha_{\rm{S}}}{\epsilon_{\rm{t}}} \frac{H_{\rm{Sun}}}{\sigma_{\rm{SB}}} \right)^{\frac{1}{4}}, \qquad H_{\rm{Sun}} = \frac{1 \ \rm{SC}}{d^2}.
 \end{equation}

\begin{figure}
  \centering
    \includegraphics[width=0.5\textwidth]{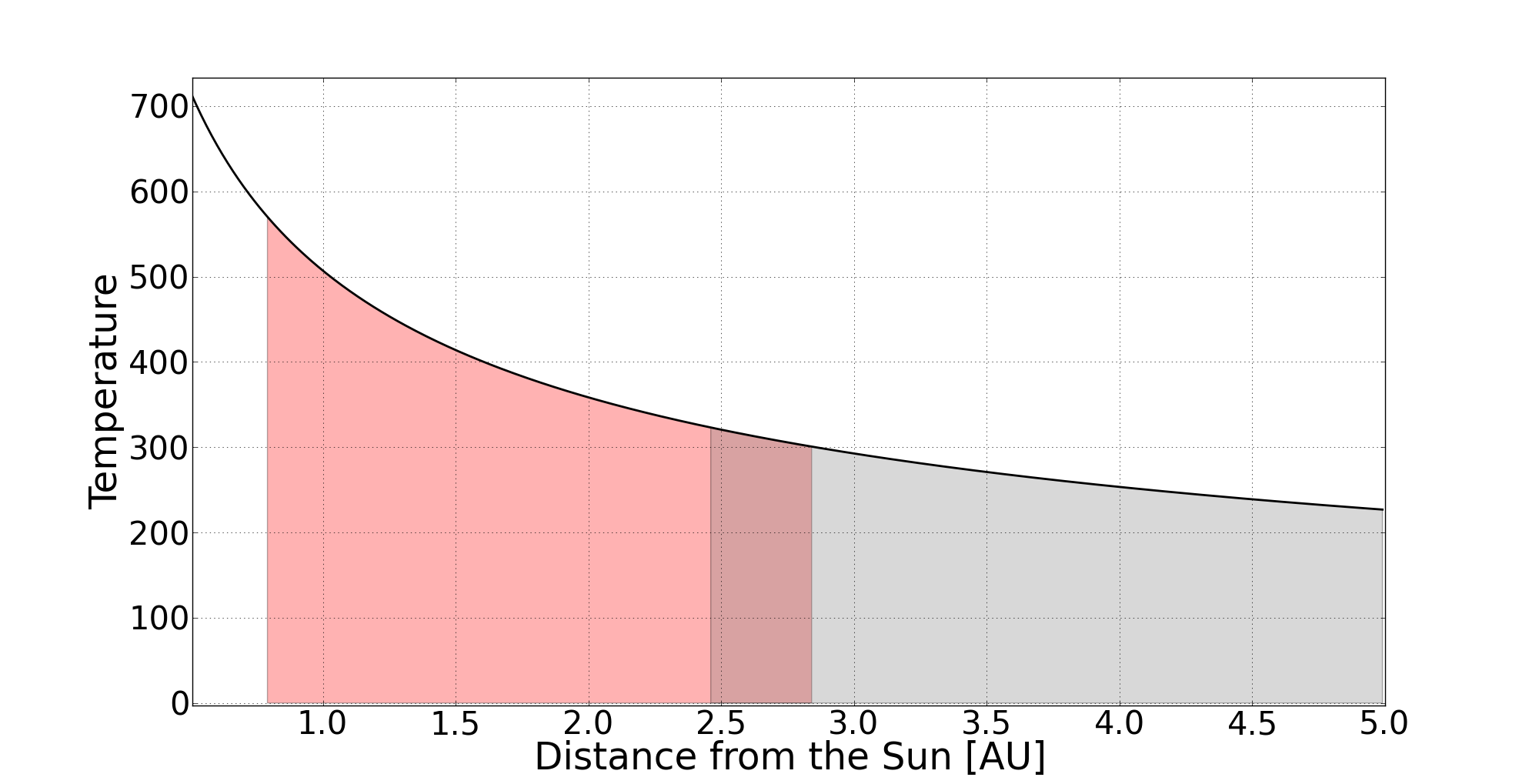}
  \caption{Temperature of the $\rm{Upilex-S^{\textregistered}}$ foil covered on both sides with $100$ nm vacuum deposited Aluminum layer as a function of the distance from the Sun. The light-red area represents temperature range in which the bubble formation was reported in the literature. The red area is the temperature range in which the formation has been confirmed by studies presented in this paper.}
  \label{UBE_temperature}
\end{figure}

\noindent Here, $A_{\rm{a}}$ is the area of the sample which absorbes the electromagnetic radiation, while $A_{\rm{e}}$ is the area which emitts the heat by radiation. Hence, the ratio $\frac{A_{\rm{a}}}{A_{\rm{e}}}$ equals $0.5$. $\sigma_{\rm{SB}}$ is the Stefan-Boltzmann constant. The thermo-optical parameters have been provided by the manufacturer of the $\rm{Upilex-S}^{\rm{\textregistered}}$ foil, the UBE company. Solar absorptance $\alpha_{\rm{S}}$ and normal emittance $\epsilon_{\rm{t}}$ are $0.093$ and $0.017$, respectively. The foil temperature as a function of distance from the Sun is represented by solid line in Fig. \ref{UBE_temperature}. Note, that the heat released by stopped protons is negligible small in comparison to the Sun's input. The light-red area (570 - 300 K) is the temperature range in which the bubble formation has been confirmed by the terrestrial laboratory experiments. Unfortunately, commonly used experimental procedures to estimate the maximum temperature at which the bubble formation is stopped, are not suitable for the real space conditions. The real critical temperature may be lower, and it has to be validated experimentally. The gray area (below 323 K) represents temperatures at which the bubble formation has been confirmed by the experimental findings presented in this paper, see Section \ref{results}. The irradiation tests have been performed for the samples' temperature of 323 K ($\sim$ 2.5 AU). Gray area represents a zone in space where bubble formation has been initiated ($\sim$ 2.5 AU). Then the bubble growth continues even when the probe is moving outwards from the Sun ($\ge$ 2.5 AU). Obviously, at larger distances the bubble growth slows down, since the probe is being bombarded by smaller proton fluxes, see Eq. \ref{flux_r}.         

%---------------------------------------------------------------
\section{Thermodynamic approach to blistering process} \label{th_model}

Here we propose a thermodynamic model of bubble formation. The model is based on the assumption that the growth proceeds quasi-static i.e. during a $\rm{j^{th}}$ period of time ${\Delta}t_{\rm{j}}$ a small portion of $\rm{H_2}$-molecules, $N_{\rm{H_2,i,j}}$, is added to the $\rm{i^{th}}$ bubble and a thermodynamic equilibrium is established rapidly.  

For simplicity it is assumed that a single bubble is a half sphere with radius $r_{\rm{i}}$. The gas within a bubble behaves to a good approximation like an ideal gas:

\begin{equation} \label{pressure}
  p_{\rm{i}}V_{\rm{i}} = \sum_{\rm{j}}^{\rm{N}}N_{\rm{H_2, i, j}} k_{\rm{B}} T,
\end{equation}

\noindent where $p_{\rm{i}}$ is the pressure of the gas, $k_{\rm{B}}$ is the Boltzmann constant, $T$ denotes the temperature of the sample, $\rm{N}$ is the number of tiny time steps, hence the total irradiation time of the sample, during which the bubble growth appears is $N \times {\Delta}t_{\rm{j}}$.

The number of recombined H atoms, subtracted by those which diffuse from the sample out (${\Delta}N_{\rm{diff,j}}$) is: 

\begin{eqnarray}
  {\Delta}N_{\rm{H,j}} &=& {\Delta}N_{\rm{p^+, j}}(1 - BS) + {\Delta}N_{\rm{diff,j}}, \\ \nonumber
  {\Delta}N_{\rm{p^+, j}} &=& I_{\rm{E}} {\Delta}t_{\rm{j}} A, \\ \nonumber
  {\Delta}N_{\rm{diff,j}} &=& - D_{\rm{H}}(T) \frac{\zeta_{\rm{H,j}}}{d_{\rm{PR}}(E)} A {\Delta}t_{\rm{j}},
\end{eqnarray}

\noindent where $A$ is the area of the sample irradiated by the protons, ${\Delta}N_{\rm{p^+, j}}$ is the number of protons sent to the sample during the $\rm{j^{th}}$ time step ${\Delta}t_{\rm{j}}$. $BS$ is the factor of backscattered ions. If $BS$ is 1 then all of the incident ions are backscattered. If $BS$ is 0 then all of the incident ions penetrate the target. $D_{\rm{H}}(T)$ is the diffusion coefficient for H atoms in a given material, $\zeta_{\rm{H,j}}$ is the number density of H atoms which may diffuse through the lattice within the $j^{\rm{th}}$ period of time, $d_{\rm{PR}}(E)$ is the so-called projected range. It is defined as an average value of the depth to which a charged particle will penetrate in the course of slowing down to rest. This depth is measured along the initial direction of the particle, and it depends on the kinetic energy of the particle \citep{pstar}. 

The number of Hydrogen molecules added during the $j^{\rm{th}}$ period of time to the $i^{\rm{th}}$ bubble $N_{\rm{H_2, i, j}}$ is then given by:

\begin{eqnarray}
  N_{\rm{H_2, i, j}} & = & 0.5 \left( N_{\rm{B}}^{\rm{T}} \right)^{-1} {\Delta}N_{\rm{H, j}} \ \eta_{\rm{max}}(s) \ \xi, \\ \nonumber
  N_{\rm{B}}^{\rm{T}} & = & N_{\rm{B}} A. \\ \nonumber
\end{eqnarray}

\noindent Here $0.5$ denotes that a single $\rm{H_2}$ molecule consists of two H atoms, $N_{\rm{B}}^{\rm{T}}$ is the total number of bubbles on the irradiated sample, $N_{\rm{B}}$ is the number of bubbles per unit area. While 100\% of protons recombine into H atoms in the metal lattice, only a part of them recombine to $\rm{H_{2}}$ molecules \citep{canham}. Hence the $\eta_{\rm{max}}(s)$ coefficient is the ratio between the number of $\rm{H_{2}}$-molecules and the H-atoms in the lattice. A $\rm{H_2}$ molecule is formed when electrons of two H atoms have anti-parallel spin $s$, otherwise the molecule cannot be created. Therefore, at most half of the H atoms can form $\rm{H_2}$ molecules, hence $\eta_{\rm{max}}(s) = 0.5$. Not all of the $\rm{H_{2}}$-molecules will merge into $\rm{H_{2}}$-clusters and finally form $\rm{H_{2}}$-bubbles. Thus, the coefficient $\xi$ denotes the ratio of the number of $\rm{H_{2}}$-molecules inside and outside the bubbles.    

The first step to estimate the radius of the $\rm{i^{th}}$ bubble is to calculate the Helmholtz free energy of the whole configuration, $F_{\rm{config}}$. Since the free energy is an additive quantity, the total free energy of bubble formation is the sum of following quantities: free energy of $\rm{H_2}$ gas inside the $\rm{i^{th}}$ bubble ($F_{\rm{gas, i}}$), of the metal surface deformation ($F_{\rm{md}, i}$) caused by the bubble growth itself, of the surface free energy ($F_{\rm{surf, i}}$) of the bubble cap, of the free energy of $\rm{H_{2}}$-molecules ($F_{\rm{H_2}}$) and of H-atoms ($ F_{\rm{H}}$) placed outside the bubbles but within the metal lattice. The Helmholtz free energy of the whole configuration described above is then: 

\begin{equation} \label{f_system}
F_{\rm{config}} = F_{\rm{gas, i}} + F_{\rm{md, i}} + F_{\rm{surf, i}} + F_{\rm{H_2}} + F_{\rm{H}}.
\end{equation}

\noindent The next step is to estimate the free energy of the $\rm{i^{th}}$ bubble. It consists of the free energy of the gas filled in the bubble, the free energy of metal deformation, and of the bubble cap surface free energy. 

Using the thermodynamic relation between gas pressure and its Helmholtz free energy $p = \left( \frac{\partial F}{\partial V} \right)_T$ together with the equation of state Eq. \ref{pressure}, the free energy of a gas within the $\rm{i^{th}}$ bubble is:

\begin{equation} \label{f_gas}
  F_{\rm{gas, i}} = - \sum_{\rm{j}}^{\rm{N}} N_{\rm{H_2, i, j}} k_{\rm{B}} T \ln \left( \frac{V_{\rm{max, i}}}{V_{\rm{min}}} \right),
\end{equation}

\noindent where $V_{\rm{max, i}}$ is the maximum volume of a given bubble. The model assumes that two $\rm{H_2}$ molecule form the smallest ("initial") possible bubble, its volume is denoted by $V_{\rm{min}}$. The radius of such a bubble is approximately $3.2$ Bohr radii \citep{ree}. Every bubble will crack if the pressure of the gas inside is higher than the pressure exerted by the metal deformation of the cap. The relation between the pressure of the gas, the strain $\sigma$ in the metal, and the bubble radius corresponding to $V_{\rm{max,i}}$ is \citep{lau}:  

\begin{equation} \label{cracking_condition}
  p_{\rm{gas, \ insite \ bubble}} - p_{\rm{outside \ bubble}} = \frac{2 \sigma} {r_{\rm{max, i}}}.
\end{equation} 

\noindent Since the sample is placed in vacuum, the pressure outside the bubble is set to zero. 

The free energy of metal deformation $F_{\rm{md}, i}$ caused by the gas pressure inside the bubble with radius $r_{\rm{i}}$ can be found in \citep{ll}, and is given by:

\begin{equation} \label{f_metal}
  F_{\rm{md, i}} = \frac{4\pi}{3} \frac{r_{\rm{i}}^3 (1 + \gamma)}{E} p_{\rm{i}}^2.
\end{equation}

\noindent Here $\gamma$ is the Poisson coefficient, i.e. ratio of transverse to axial strain of a sample material, $E$ is the Young's module.  

The free energy of a surface of a cap of the $\rm{i^{th}}$ bubble is given by \citep{marty}:

\begin{equation} \label{f_surface}
F_{\rm{surf, i}} = 4 \pi r_{\rm{i}}^2 \sigma(T).
\end{equation}

The Helmholtz free energy of the $\rm{H_{2}}$-molecules located at certain positions in the metal lattice but outside the bubbles is calculated in the form $F = E_{\rm{int}} - TS$. Where $E_{\rm{int}}$ is the internal energy of molecules/atoms located at certain positions in the metal lattice. Applying the statistical definition of the entropy $S$, this free energy is: 

\begin{eqnarray} \label{f_h2}
  F_{\rm{H_2}} &=& \left( N_{\rm{H_{2}}}^{\rm{T}} - \sum_{\rm{i}}^{N_{\rm{B}}^{\rm{T}}} \sum_{\rm{j}}^{\rm{N}} N_{\rm{H_{2}, i, j}} \right) \\ \nonumber
 & \times & \left[ \epsilon_{\rm{H_2}} + k_{\rm{B}}T \ln \left( \frac{N_{\rm{H_2}}^{\rm{T}} - \sum_{\rm{i}}^{N_{\rm{B}}^{\rm{T}}} \sum_{\rm{j}}^{\rm{N}} N_{\rm{H_{2}, i, j}}}{N_{\rm{0}}} \right) \right], 
\end{eqnarray}

\noindent where $N_{\rm{H_{2}}}^T$ is the total number of $\rm{H_2}$ molecules inside the sample, $\epsilon_{\rm{H_2}}$ is the binding energy of $\rm{H_{2}}$ molecule to a vacancy. $N_{\rm{0}}$ is the number of lattice sites, which can be expressed by:

\begin{equation} \label{lattice_sites}
  N_{\rm{0}} = N_{\rm{A}}d_{\rm{PR}}\frac{A}{M_{\rm{u}}},
\end{equation}

\noindent where $N_{\rm{A}}$ is the Avogadro's number. $M_{\rm{u}}$ is the molar mass of the sample's material.

The Helmholtz free energy of H atoms located at certain positions within the metal lattice is:

\begin{equation} \label{f_h}
  F_{\rm{H}} = \left(N_{\rm{H}}^{\rm{T}} - 2N_{\rm{H_{2}}}^{\rm{T}}\right) \left[ \epsilon_{\rm{H}} + k_{\rm{B}}T \ln \left( \frac{N_{\rm{H}}^{\rm{T}} - 2N_{\rm{H_{2}}}^{\rm{T}}}{N_{\rm{0}}}\right) \right],
\end{equation}

\noindent where $\epsilon_{\rm{H}}$ is the migration energy of the H atom in the metal lattice, and $N_{\rm{H}}^{\rm{T}}$ is the total number of H atoms in the sample. 

Since now each term of the Eq. \ref{f_system} is determined, the next step is to estimate the radius $r_{\rm{i}}$ of the $\rm{i^{th}}$ bubble at given time $t$. This will be achieved by assuming that the process of bubble growth is quasi-static, i.e. during each $\rm{j^{th}}$ time step ${\Delta}t_{\rm{j}}$ a small portion of $\rm{H_2}$ molecules is merged to the $\rm{i^{th}}$ bubble and the thermodynamic equilibrium is rapidly re-established:

\begin{equation} \label{condition}
  \frac{\partial F_{\rm{system}}}{\partial N_{\rm{H_2, i, j}}} = 0.
\end{equation}

\noindent This condition leads to the following fifth order equation for $r_{\rm{i}}$:

\footnotesize
\begin{eqnarray} \label{bfc} 
 &8 \pi \Xi_{\rm{i,j}} \sigma(T) r_{\rm{i}}^5 - H_{\rm{i}}r_{\rm{i}}^4 + \frac{3}{\pi} \frac{1+\gamma}{E} \left(\sum_{\rm{j}}^{\rm{N}} N_{\rm{H_2, i, j}} \right) k_{\rm{B}}^2 T^2 \\ \nonumber
& \times \left[ 2 N r_{\rm{i}} - 3 \Xi_{\rm{i,j}} \sum_{\rm{j}}^{\rm{N}} N_{\rm{H_2, i, j}} \right] = 0,
\end{eqnarray}
\normalsize

\noindent $\Xi_{\rm{i,j}}$ is defined below in Eq. \ref{dyn_model}, $H_{\rm{i}}$ denotes the abbreviation:

\begin{equation}
H_{\rm{i}} = - \frac{\partial F_{\rm{gas, i}}}{\partial N_{\rm{H_2, i, j}}} - \frac{\partial F_{\rm{H}}}{\partial N_{\rm{H_2, i, j}}} - \frac{\partial F_{\rm{H_2}}}{\partial N_{\rm{H_2, i, j}}}. 
\end{equation}

A realistic model of bubble radius growth, $\Xi(i,j)$, can be estimated by following Gedankenexperiment. Obviously at the beginning of the bubble growth process, the differential increase of the bubble radius is higher than at its end. It is implied, that the number of $\rm{H_2}$ molecules in the system is conserved and at each time step one of them merges into a bubble. After $\Delta{t}$ the bubble consists of $2\rm{H_2}$ molecules, hence the number of molecules increases by $50\%$. At the time $2\Delta{t}$ the bubble consists of $3\rm{H_2}$ molecules, hence the number increase is now $33.3\%$, and so on. Therefore $\Xi$ is: 

\begin{equation} \label{dyn_model}
  \Xi_{\rm{i,j}} = \frac{\Delta r_{\rm{i}}}{\Delta N_{\rm{H_2,i,j}}} = j^{\alpha} r_{\rm{i, 0}}, \qquad \alpha = \frac{1}{3} 
\end{equation}  

\noindent The exponent $\alpha$ is a model parameter of the bubble growth. The value $\frac{1}{3}$ corresponds to the Gedankenexperiment presented above. However, the true value of the $\alpha$ parameter differs from that . In the process of bubble growth, particles (the Hydrogen) are added to the system i.e. the probe is permanently irradiated by the protons, they penetrate the target and recombine to the Hydrogen. On the other hand, both, due to the diffusion process and bubble cracking, some Hydrogen atoms leave the system. Therefore, the number of Hydrogen atoms in the system is not conserved. Hence, a series of experiments have been performed to estimate a realistic $\alpha$ parameter; results are presented in Section \ref{results}.

\subsection{The effect of bubble formation onto the specular reflectivity} \label{spec_reflectivity}

The momentum transfer of a photon to an ideal reflecting surface is given by $\Delta{q}=2q\cos \theta$, where the factor 2 is just in accordance with specular reflectivity. Certainly, the surface quality will suffer during the irradiation with protons from progressing bubble formation. At time $t=0$ the foil has not been exposed to the electromagnetic radiation and/or charged particles, and is considered to be a perfect mirror with the reflectivity of $R=1$. It means that all of the incident light rays are reflected perfectly, no light ray is absorbed or diffusively reflected by the target. Later, when the foil has been irradiated by a flux of protons and molecular Hydrogen bubbles have been formed on its surface, the reflectivity of the degraded foil will be reduced. This deterioration is calculated in the following way: the foil is covered by a grid with a fixed single cell size of $\epsilon \times \epsilon$. The reflectivity of a single cell is by definition $\frac{\Delta q}{\Delta q_{\rm{max}}}$, where ${\Delta}q$ is momentum transfer of a photon to the $\rm{i^{th}}$ cell of the degraded foil, while ${\Delta}q_{\rm{max,i}}$ is the momentum transfer of a photon to the $\rm{i^{th}}$ cell of a perfect mirror. 

Therefore, taking into account all cells, one has:

\begin{equation} \label{reflectivity_common}
\Delta R = \frac{\sum_{\rm{i}}^{\rm{N_{cell}}} \Delta q_{\rm{i}} }{ \sum_{\rm{i}}^{\rm{N_{cell}}} \Delta q_{\rm{max, i}} }.
\end{equation}

\noindent Here $N_{\rm{cell}}$ is the number of cells. The path of photons is directed parallel to the foil surface normal. Therefore, at time $t=0$ the foil was a perfect mirror without surface imperfections and $\theta_{\rm{i}}=0$. Later, when the surface is populated with bubbles, $\theta_{\rm{i}}$ will vary between $0^o$ and $90^o$. Thus, Eq. \ref{reflectivity_common} reduces to: 

 \begin{equation}
\Delta R = \frac{\sum_{\rm{i}}^{\rm{N_{cell}}} 2q \cos \theta_{\rm{i}} }{ N_{\rm{cell}} \times 2q } =  \frac{\sum_{\rm{i}}^{\rm{N_{cell}}} \cos \theta_{\rm{i}} }{ N_{\rm{cell}} }.
\end{equation}

Change of the foil's reflectivity due to growing population of $\rm{H_2}$-bubbles is presented in Section \ref{results}.

%----------------------------------
\section{Results} \label{results}

\begin{figure}
  \centering
   \includegraphics[width=0.4\textwidth]{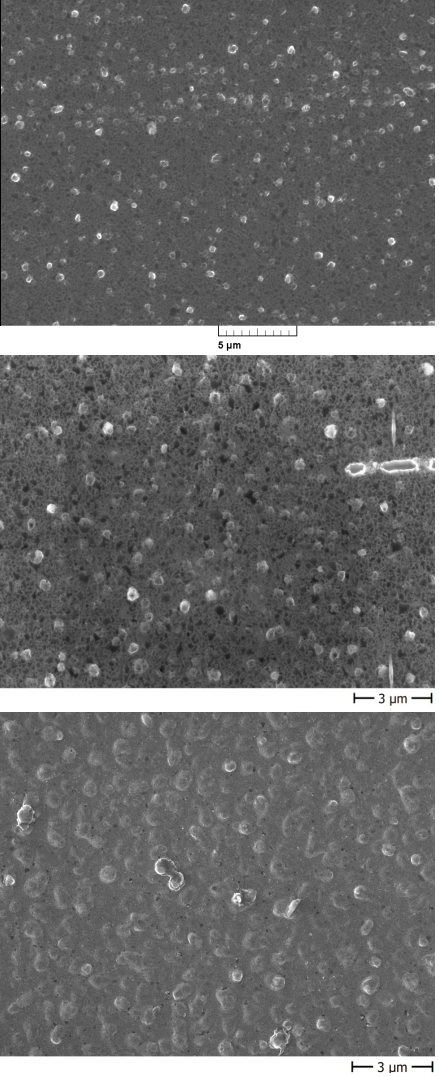}
   \caption{Electron microscope pictures of probes A1 (top), A2 (middle), and A3 (bottom).}
  \label{dose_dep}
\end{figure}

To validate the model the following experiments were performed. Three probes (A1, A2, and A3) were exposed to a flux of 2.5 keV protons, each one with longer irradiation time, see Table \ref{test_parameters}, where $t_{\rm{S}}$ is a time in space until a probe will collect a given dose of protons. Results are shown in Fig. \ref{dose_dep}. From top to bottom, the pictures correspond to the probes A1, A2, and A3, respectively. Average sizes of bubbles have been estimated to $0.17 \ \pm 0.05$ $\rm{\mu}$m, $0.2 \ \pm 0.05$ $\rm{\mu}$m, and $0.25 \ \pm 0.05$ $\rm{\mu}$m for probe A1, A2, and A3, respectively. There is a strict correlation between a dose of protons and the average bubble size for a given population. The higher the proton dose, the larger the bubble sizes. Examining the electron microscope pictures, the surface density of bubbles has been estimated to $\sim 10^8$ $\rm{cm^{-2}}$. 

\begin{table}[!ht]
\centering 
\caption{Test parameters}
\scalebox{0.85}{\begin{tabular}{ccccc}
\hline
Probe symbol & $T$ [K] & $E$ [keV] & $D$ [$\rm{p^+ \ cm^{-2}}$] & $t_{\rm{S}}$ [days]  \\
\hline
A1 & 323.0 & 2.5 & $7.8 \times 10^{17}$ & 4.8\\
A2 & 323.0 & 2.5 & $8.2 \times 10^{17}$ & 5.0\\
A3 & 323.0 & 2.5 & $1.3 \times 10^{18}$ & 7.9\\
\hline
\end{tabular}}
\label{test_parameters}
\end{table}

For numerical simulation a $10 {\mu}m \times 10 {\mu}m$ foil was specified. That choice allows to simulate a smaller number of bubbles, i.e. it decreases the computation time of the simulation. It implies also an important assumption that surface arrangement of the bubbles is isotropic i.e. any $10 {\mu}m \times 10 {\mu}m$ area of the irradiated sample is indistinguishable. Table \ref{model_parameters} collects all of the model parameters used in the simulation. The first set of parameters characterize mechanical and thermo-optical properties of vacuum deposited Aluminum on UBS's $\rm{Upilex-S^{\textregistered}}$ foil. Second set specifies values of the parameters which have been used to fit the model to the experimental data presented here.

\begin{figure}
  \centering
   \includegraphics[width=0.36\textwidth]{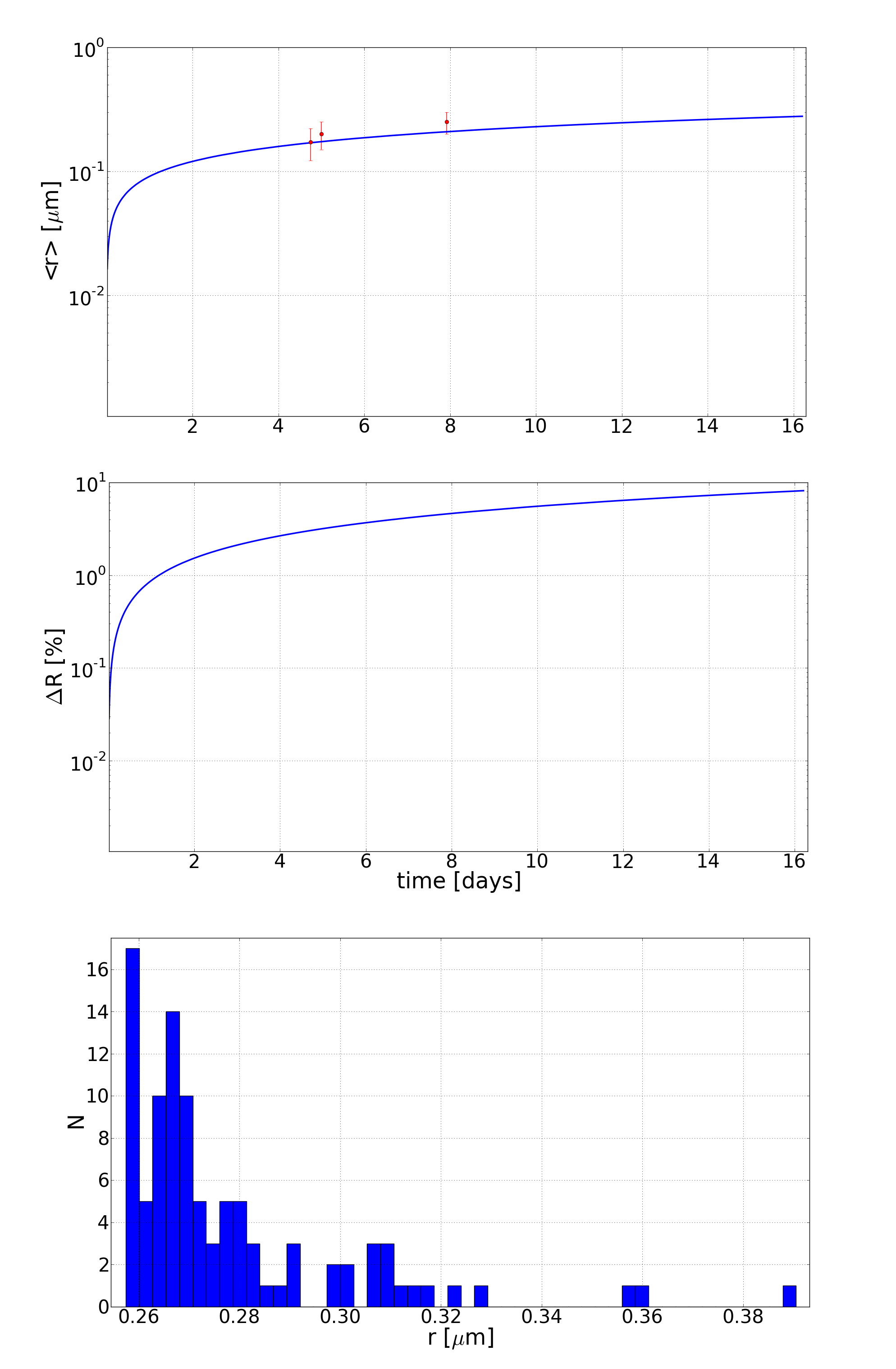}
   \caption{Time evolution of an average bubble radius from the population (top plot), specular reflectivity decrease due to bubble growth (middle plot), bubble size distribution at a $100$ $\rm{{\mu}m^2}$ at the end of the simulation (bottom plot).}
  \label{model}
\end{figure}

To fit the proper gradient of bubble growth, the $\alpha$ parameter was set to $0.6$, see Eq. \ref{dyn_model}. Comparison of the average bubble size of the experimental and numerical findings are drawn in the top plot of the Fig. \ref{model}. The $\xi$ parameter was set to $0.98$. It determines the height of the curve. A decrease of the specular reflectivity of the foil as a function of time is shown in the middle plot of the Fig. \ref{model}. The decrease of the reflectivity is $3.0$, $3.2$, and $4.6$ \% in comparison to the non-irradiated foil for $4.75$, $5.0$, and $7.9$ $t_{\rm{S}}$, respectively. At the end of the simulation, the decrease of the reflectivity is $8$ \%. Clearly, the larger the bubble sizes, the larger the specular reflectivity decrease ${\Delta}R$ in comparison to the non-irradiated foil. A distribution of the bubbles at the end of the simulation ($t_{\rm{S}} = 16$ days) is drawn in the bottom plot of the Fig. \ref{model}. Most of the bubbles have a typical size in range $0.26$ to $0.28$ $\rm{\mu}$m, there are only a few which have sizes larger than $0.3$ $\rm{\mu}$m.      

\begin{table}[!ht]
\centering 
\caption{Model parameters}
\scalebox{0.75}{\begin{tabular}{ccl}
\hline
Symbol & Value & Description  \\
\hline
$\varrho$ & $2.7$ [$\rm{g \ cm^{-3}}$] & Al density \\
 $M$       & $26.98$ [$\rm{g \ mol^{-1}}$] & Al molar mass \\
$E$       & $69 \times 10^{10}$ [$\rm{dyn \ cm^{-2}}$] & Al Young modulus \\
$\gamma$  & $0.33$ & Al Poisson coefficient \\
 $\epsilon_{\rm{H}}$ & 0.52 [eV] & H migration energy in the Al lattice \citep{lind} \\
$\epsilon_{\rm{H_2}}$ & 0.06 [eV] & $\rm{H_2}$ binding energy to a vacancy in Al \citep{lu} \\ 
$\alpha_{\rm{S}}$ & 0.093 & solar absorptance \\
$\epsilon_{\rm{t}}$ & 0.017 & normal emittance \\
\hline
$BS$ & 0.02 & $\rm{H^+}$ back scattering factor \citep{srim} \\
$A$ & $\rm{10 {\mu}m \times 10 {\mu}m}$  & irradiated area\\
 $T$ & 323 [K] & sample's temperature\\
$\eta_{\rm{max}}(s)$ & 0.5 & $\rm{\frac{H_2 \ lattice}{H \ lattice}}$ ratio \\  
$\xi$ & 0.98 & $\frac{\rm{H_2 \ bubbles}}{\rm{H_2 \ lattice}}$ ratio\\
 $\alpha$ & 0.6 & bubble growth parameter \\
 $N_{\rm{B}}$ & $10^8$ $\rm{cm^{-2}}$ & number of bubbles per unit area\\
\hline
\end{tabular}}
\label{model_parameters}
\end{table}

%----------------------
\section{Conclusions} \label{conclusions}

It has been proven that thermodynamic model is a flexible tool to simulate and to reproduce the real growth of the molecular Hydrogen bubbles. However, the estimated  $\alpha$ and $\xi$ parameters are applicable only for the here presented experimental findings. These parameters depend on type and temperature of the irradiated material. Therefore, change of the material type and the experimental conditions requires a new validation of the model. 

The time evolution of decrease of the specular reflectivity ${\Delta}R$ is a model prediction. It is highly required to perform experimental confirmation of that findings, since the real reflectivity decrease can differ from that estimated here.

The thermodynamic model requires further improvements. The considered aging factor, the solar protons, are not the only one which can influence the bubble growth process. The solar wind is also essentially made up of electrons and small proportion of heavier ions \citep{meyer}. Additionally, electromagnetic radiation in the lower wavelength range has to be taken into consideration. These degradation factors can slow down the bubble growth. The growth deceleration can be explained as follows. Hydrogen molecule dissociate at the energy of $4.5$ eV \citep{balak}. The dissociation may be caused by the UV-light at wavelengths $\le 274$ nm. The $\rm{H_2}$ gas within the bubbles can then be partially dissociated, and H atoms can diffuse easily through the bubble caps. As a result the bubble growth process may slow down. The deceleration can be strengthened by heavier ions generated by the Sun e.g. $\alpha$-particles. Their diameter is much larger than that of protons or electrons, hence, collisions between the $\rm{H_2}$ molecules and the $\alpha$-particles within the bubbles can additionally increase the dissociation efficiency.

The present condition for the bubble crack mechanism, Eq. \ref{cracking_condition}, assumes that the pressure outside the bubbles is negligible small. Under the real space conditions the electromagnetic radiation will exert a pressure on the caps, hence, their sizes may be smaller. On the other hand, bubble caps loose thermal contact with the base material and they become overheated \citep{astrelin}. As a consequence the caps can brake and the $\rm{H_2}$ gas can be released. That aspect of the blistering process needs to be examined.

By these reasons further experimental studies are planned. First, the temperature range in which the bubble formation takes place has to be evaluated. Afterwards, the mentioned influence of the UV-light on the bubble growth dynamics will be studied. 

\section*{Acknowledgments}
We would like to express our special thanks to dr. Herbert Juling who performed electron microscope measurements.

\end{document}